\documentclass{kapproc} 

\usepackage{procps} 

\usepackage[dvips]{graphicx}

\upperandlowercase

\kluwerbib 

\begin{document}

\articletitle
{Spectral analysis of 50 GRBs detected by HETE-2}

\author{C. Barraud and the HETE-2 team}

\affil{Observatoire Midi-Pyrenees, 14 av. E. Belin, 31400 Toulouse, France}
\email{barraud@ast.obs-mip.fr}

\begin{abstract}

FREGATE, the gamma-ray detector of HETE-2 is entirely dedicated to the study of GRBs. 
Its main characteristic is its broad energy range, from 7 keV to 400 keV. This energy range 
can be further extended down to 2 keV using the data from the WXM, the X-ray detector of HETE-2.
 Such a large energy range allows to study in details the prompt emission of GRBs, determining 
with a high precision their spectral parameters. Moreover, because this energy range extends at low energies, the sample of GRBs
detected by both  FREGATE and WXM contains a significant fraction of X-Ray Rich GRBs and X-Ray Flashes. 

We present here the distributions of the spectral parameters mesured for the time integrated 
spectra of 50 GRBs. We  put emphasis on the distribution of the low energy spectral index $\alpha$. Because 
FREGATE and WXM detected all classes of GRBs, we also discuss the connection between GRBs, X-Ray Rich GRBs 
and X-Ray Flashes.

\end{abstract}



\section{Introduction}
\begin{figure}
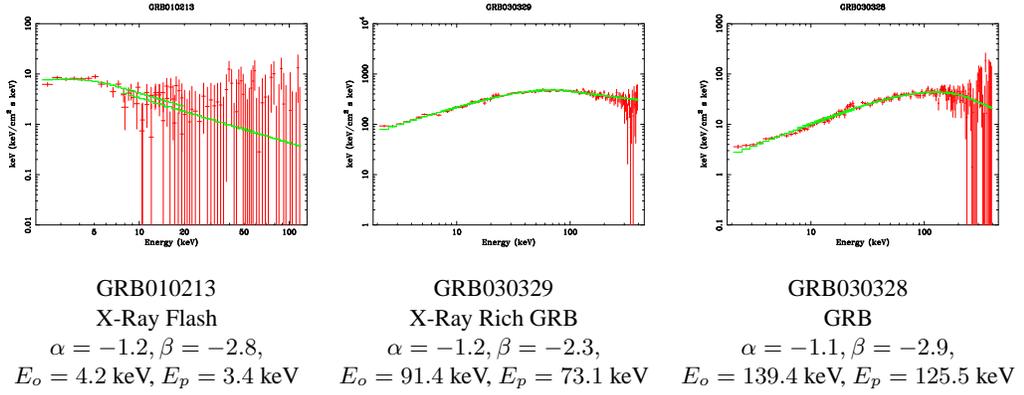

\begin{tabular}{ccc}
\includegraphics[angle=-90,width=4cm]{GRB010213grbm.ps} &\includegraphics[angle=-90,width=4cm]{GRB030329grbm.ps}&\includegraphics[angle=-90,width=4cm]{GRB030328grbm.ps} \\
&&\\
GRB010213 & GRB030329 & GRB030328 \\
X-Ray Flash & X-Ray Rich GRB & GRB \\
$\alpha = -1.2 , \beta = -2.8,$ & $\alpha = -1.2 , \beta = -2.3, $ & $\alpha = -1.1 , \beta = -2.9, $\\
$E_o = 4.2$ keV, $E_p = 3.4$ keV & $E_o = 91.4$ keV, $E_p = 73.1$ keV & $E_o = 139.4$ keV, $E_p = 125.5$ keV \\ 
\end{tabular}
\caption{Spectra of the different classes of GRBs: X-Ray Flashe, X-Ray Rich GRB and GRB using both FREGATE and WXM data.}
\label{spectre}
\end{figure}
FREGATE is the gamma-ray detector of HETE-2 (see \cite{atteia03} for a description of FREGATE).
Its broad-energy range 7--400 keV which can be extended down to 2 keV using the WXM instrument (see
 \cite{wxm} for a description of WXM) allows us to determine with high precision 
the spectral parameters of the prompt emission of the GRBs seen by both instruments. The two instruments 
also detected an important fraction of X-Ray Rich GRBs and X-Ray Flashes (see \cite{heise01} for a  description of these new classes) and we are now able to discuss 
the differences and the similarities between these three populations.

We present in this paper an update of the results presented in \cite{barraud03}: 'Spectral analysis of 
35 GRBs/XRFs observed with  HETE-2/FRE\-GATE'. This paper presented a first spectral analysis 
of 35 GRBs detected by HETE-2/FREGATE since its launch in October 2000 and which were well localized by either the 
instruments on-board HETE-2 (WXM or SXC, see \cite{sxc} for a description of the Soft X-Ray Camera), 
 or by the GRB InterPlanetary Network (IPN). 

 The update of the paper \cite{barraud03} corresponds to an increase of the number of GRBs 
seen by both FREGATE and WXM which now reaches 50, all in the class of long GRBs. We didn't include 
the two short/hard bursts GRB020113 and GRB020531 detected by HETE-2. 
Another improvement is that the spectral parameters are now obtained from a joint fit of WXM and FREGATE data. We thus  obtain spectra 
ranging from 2 keV to 400 keV.

We focus here on the distribution of the spectral parameters: we show that the distribution of the low energy spectral index
 $\alpha$ is compatible with the predictions of the synchrotron shock model and we show that a
 significant fraction of bursts have their peak energy $E_p$ lower than 50 keV.
We also put emphasis on the hardness-intensity correlation. This correlation shows that the three populations, GRBs, X-Ray-Rich
GRBs and X-Ray-Flashes form a continuum which strongly suggest that they are all produced by the same phenomenon. 

\section{The spectral analysis}

Our sample is made of 50 GRBs localized either with the HETE-instrument  or with the GRB Interplanetary Network
and which were within $60^o$ of the FREGATE line of sight.
GRB spectra are usually fit with  the BAND function (\cite{band93}), which is two power laws smoothly connected:

\begin{tabular}{ll} \label{band}
 $N(E) = A E^{\alpha} exp(\frac{-E}{E_o})$ &   for $E > (\alpha - \beta) E_o$,\\
 $N(E) = B E^{\beta}  $ &                     otherwise.\\
\end{tabular}
\\In this equation $\alpha$ and $\beta$ are the photon indices of respectively the low and the high energy power laws, $E_o$ is the energy break  and the peak 
energy $E_p$ of the $\nu f_{\nu}$ spectrum is defined by : $Ep= Eo*(2+\alpha)$. 
We have to notice that in the case of GRBs detected by HETE-2, the energy range is often not broad 
enough to determine accurately 
all the parameters of the spectra especially the index of the high energy power law $\beta$. 
In these cases, and in order to not neglect the flux at high energies,  we fix the value of $\beta$ to an arbitrary value which is $-2.3$.
The combination of WXM and FREGATE data allows us to study spectra down to 2 keV and 
determine accurately the parameter $E_p$, even at low energies for the X-Ray-Rich GRBs and X-Ray-Flashes. 

Figure \ref{spectre} shows the $\nu f_{\nu}$ spectrum of one GRB in each of the three classes derived 
from joint fits of the 
WXM and FREGATE data.
The left panel shows the first  X-Ray-Flash detected by HETE-2, GRB010213. The addition of 
the WXM data allowed to determine the  $E_p$ which has a value of 3.4 keV. This is the 
weakest GRB detected by FREGATE. The middle panel is GRB030329 the ``monster burst'', an X-Ray Rich with $E_p=73.1$ keV, and the right panel is GRB030328 a ``standard'' GRB with $E_p = 125.5$ keV.

\section{The distribution of the spectral parameters}

\begin{center}
\begin{figure}
\includegraphics[width=7cm]{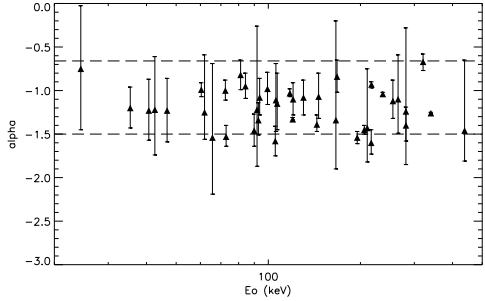}
\label{alphaeo}
\vspace{-1.cm}
\caption{The photon index of the low energy power law $\alpha$ versus the break energy $E_o$,}
\end{figure}
\end{center}

Figure \ref{alphaeo} displays $\alpha $, the photon index of the low energy power law 
versus $E_o$, the energy break for 41 GRBs for which we were able to mesure these parameters. 
For clarity of the figure, the $90 \%$  error bars are shown for $\alpha$ only. The dotted 
lines represent the limits predicted by the classical synchrotron shock model which are $-3/2$ 
and $-2/3$. The values used in this plot result from a fit of the time-integrated spectra with a cutoff power 
law model. This model is similar to the Band model but it uses only the low energy part and the spectral break
 of the band function. The definition of $E_p$ is not affected by the choice of this model. We use this procedure because in most cases the energy range of HETE--2 (2--400 keV) 
is not broad enough to determine good values of $\beta$ and the values of $\alpha$ and $E_o$ 
are less constrained if we use the Band function \ref{band}.

This figure also shows that whatever the value of Eo, all values of $\alpha$ are compatible with the 
values expected from the synchrotron shock model. In this model, the emission comes from synchrotron 
radiation emitted 
by a population of shock accelerated electrons (\cite{katz}, \cite{cohen}, \cite{llyod00}). 
We can also notice that there is a significant fraction of GRBs with $E_o$ lower than 50 keV.

\begin{center}
\begin{figure}
\includegraphics[width=7cm]{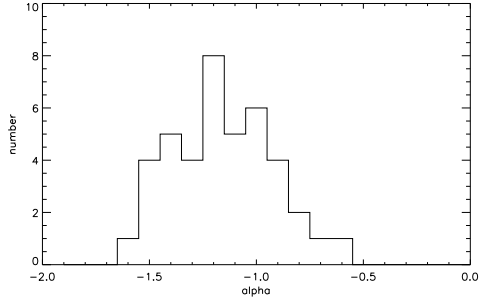}
\caption{Distribution of the spectral index $\alpha$}
\label{alphabeta}
\end{figure}
\end{center}
\vspace{-1.cm}
The histogram \ref{alphabeta} displays the distribution of the photon index of the low 
energy power law $\alpha$. This distribution peaks at $-1.2$  and has a full width at half maximum of approximately 0.5.
\section{The hardness-intensity correlation and the connection between GRBs, X-Ray-Rich
GRBs, and X-Ray-Flashes}
\begin{figure}
\includegraphics[width=7cm]{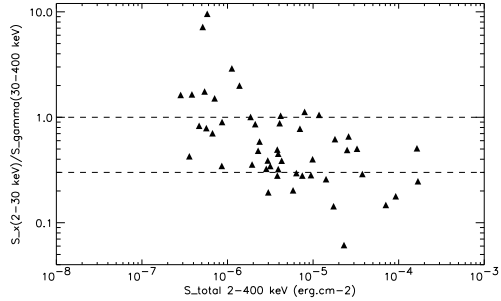}
\label{durete2}
\caption{softness ($S_x/S_\gamma$) versus total fluence for GRBs observed by HETE-2.}
\end{figure}

\begin{figure}[!h]
\includegraphics[width=7cm]{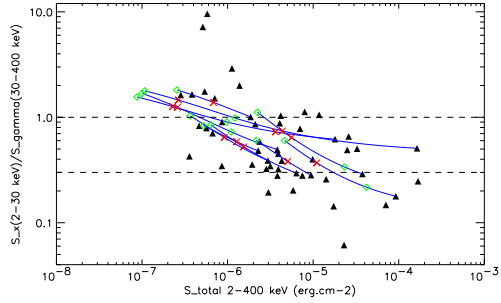}
\caption{softness ($S_x/S_\gamma$) versus total fluence for GRBs observed by HETE-2.
The lines indicates how GRBs with known redshift would evolve on this diagram if their redshifts
 were increased to z = 10.}
\label{durete1}
\end{figure}

Figure \ref{durete2} shows the hardness-intensity correlation observed by HETE-2. 
The y-axis shows the inverse of the hardness or the \emph{softness} ($S_x/S_\gamma$) which 
is the ratio between the fluence in 2--30 keV ($S_x$) and the fluence in 30--400 keV ($S_\gamma$).
 The x-axis shows the intensity, the fluence in 2-400 keV.
The first point highlighted by this figure is the strong correlation between 
these two quantities over 3 orders of magnitude in fluence, it shows that the weaker a burst is, the softer it is.
\\
The second point is that this figure does not clearly separate X-Ray-Rich GRBs 
and X-Ray-Flashes from GRBs. We define here X-Ray-Rich GRB as GRBs which have a softness 
in the range 0.3--1, and X-Ray-Flashes as GRBs which have a softness greater than 1. 
The two dotted lines represent the limits (in terms of $S_x/S_\gamma$)
 of the 3 populations. It is clear that there is no gap between these populations and 
the continuum strongly suggest that these three types of bursts are all produced 
from the same phenomenon.
\\
 We now discuss if X-Ray-Rich GRBs and X-Ray-Flashes can be highly redshifted GRBs, which 
was one of the first hypothesis to explain such weak and soft bursts. To this end we looked how 
GRBs with known redshift would evolve on this diagram (\ref{durete2}) if their redshifts
 were increased to z = 10. We added lines on figure \ref{durete1} which indicate 
the evolution of these GRBs (ie GRB010921 z = .45, GRB020124 z = 3.2, GRB020813 z = 1.25, 
GRB021004 z = 2.31, 
GRB021211 z = 1.01, GRB030226 z = 1.98, GRB030323 z = 3.37, GRB030328 z = 1.52, 
GRB030329 z = .17, GRB030429 z = 2.65). 
Redshifts 1 and 5 are marked with crosses and redshifts 2 
and 10 with empty squares. 
What we notice here is that these GRBs have their total fluence which decreases while their 
softness increases with the redshift.
We notice that the higher value of softness we reach at z = 10 with this method is $S_x/S_\gamma =2$. This value is very small compared to the $S_x/S_\gamma = 10$ found for two bursts.  This mechanism, putting GRBs at high redshift, can 
apparently produce X-Ray Rich GRBs and X-Ray-Flashes but it seems to reach an upper limit and can't produce 
the very high values of the softness observed for  X-Ray-Flashes. 
\\

\begin{figure}[!h]
\includegraphics[width=7cm]{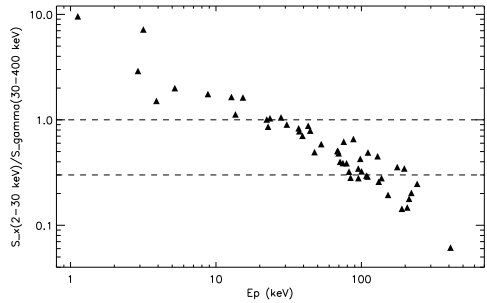}
\label{dureteep}
\caption{softness ($S_x/S_\gamma$) versus  $E_p$ for GRBs observed by HETE-2.}
\end{figure}
Figure \ref{dureteep} shows the softness versus the value of the peak energy $E_p$ 
in keV for all the GRBs detected by HETE-2.  The horizontal dashed lines represent the limits of the three classes.
 This figure first indicates that the softness is very well  representative of the value of the $E_p$. 
This is very important for spectral analysis indeed the $E_p$ is often hard to calculate because 
the value of $\alpha$ and $E_o$ are sometimes not well constrained, especially for soft bursts 
which have their value of $E_o$ near the lower limit of the energy range of HETE-2 
(for example GRB010213 has $E_o = 4.2$ keV and it is clear that the value of $\alpha$ can't be 
well determinated and so the value of $E_p$) whereas the fluence in all energy ranges can 
always be calculated.
     This plot also shows that the distribution of $E_p$ covers a broad energy range similar to that  
covered by HETE-2 from few keV to several hundred keV. In addition of the distribution 
found by BATSE which peaks at 200 keV HETE-2 had allowed the detection of many bursts which 
have their $E_p$ lower than 50 keV. This makes this distribution very broad. GRBs which have 
a low $E_p$ are associated with X-Ray-Rich GRBs (the middle part of the plot) 
and X-Ray-Flashes (the upper part of the plot).
\section{Conclusion}
In this paper, we update the results presented in \cite{barraud03}. The update consists of 
an increase of the number of GRBs to 50, and an  analysis which is now based on joint spectra with 
both WXM and FREGATE data. Joint spectra allow to study spectra from 2 keV to 400 keV and thus provide 
 more accurate values of $\alpha$, $E_o$ and fluence ratios. 
\\
The first important result of this study comes from the distribution of the spectral parameter $\alpha$ which is fully 
in agreement with the predictions of the synchrotron shock model.
\\
We have also shown that the new class of ``soft'' GRBs cannot apparently be explained as high redshift GRBs. But 
we have confirmed and extended the hardness-intensity correlation which strongly 
suggests that the three classes of GRBs, X-Ray-Rich GRBs and X-Ray-Flashes, which distinguish 
themselves by the values of their $E_p$ and their softness are all from the same phenomenon.
More GRBs and  more broad energy coverage of GRB missions will allow to refine these 
results and constrain models of the prompt emission of GRBs.


\end{document}